\documentclass[twocolumn]{aastex63}

\usepackage[utf8]{inputenc}  
\usepackage[T1]{fontenc}
\usepackage{graphicx,psfrag}
\usepackage{mathrsfs}
\usepackage{amsmath,amsfonts,amssymb,pifont}
\usepackage{multirow,enumerate}
\usepackage{comment,hyperref}
\usepackage{xcolor}
\usepackage{acronym}
\usepackage{xspace}
\usepackage{lineno}
\usepackage[normalem]{ulem}
\usepackage{mathtools}
\usepackage{subfigure}
\usepackage{makecell}

\newcommand{\ROnePointFourNICERRiley}{11.56^{+0.79}_{-0.76} \rm{km}}
\newcommand{\ROnePointFourNICERMiller}{11.62^{+0.85}_{-0.79} \rm{km}}
\newcommand{\ROnePointFourNICERBoth}{11.59^{+0.83}_{-0.76} \rm{km}}

\newcommand{\ROnePointFourNICERXMMRiley}{11.84^{+0.79}_{-0.80} \rm{km}}
\newcommand{\ROnePointFourNICERXMMMiller}{12.03^{+0.77}_{-0.87} \rm{km}}
\newcommand{\ROnePointFourNICERXMMBoth}{11.94^{+0.76}_{-0.87} \rm{km}}

\newcommand{\MmaxBoundNICERRiley}{2.15^{+0.14}_{-0.12} M_{\odot}}
\newcommand{\MmaxBoundNICERMiller}{2.16^{+0.15}_{-0.12} M_{\odot}}
\newcommand{\MmaxBoundNICERBoth}{2.15^{+0.15}_{-0.12} M_{\odot}}

\newcommand{\MmaxBoundNICERXMMRiley}{2.17^{+0.15}_{-0.13} M_{\odot}}
\newcommand{\MmaxBoundNICERXMMMiller}{2.18^{+0.15}_{-0.15} M_{\odot}}
\newcommand{\MmaxBoundNICERXMMBoth}{2.18^{+0.16}_{-0.13} M_{\odot}}

\newcommand{\MmaxNoBoundNICERRiley}{2.23^{+0.31}_{-0.20} M_{\odot}}
\newcommand{\MmaxNoBoundNICERMiller}{2.26^{+0.36}_{-0.24} M_{\odot}}

\newcommand{\MmaxNoBoundNICERXMMRiley}{2.31^{+0.37}_{-0.25} M_{\odot}}
\newcommand{\MmaxNoBoundNICERXMMMiller}{2.40^{+0.35}_{-0.32} M_{\odot}}

\newcommand{\BayesNMMA}{0.27 \pm 0.01}
\newcommand{\BayesNICERRiley}{0.30 \pm 0.01}
\newcommand{\BayesNICERMiller}{0.29 \pm 0.01}
\newcommand{\BayesNICERBoth}{0.29 \pm 0.01}
\newcommand{\BayesNICERXMMRiley}{0.23 \pm 0.01}
\newcommand{\BayesNICERXMMMiller}{0.21 \pm 0.01}
\newcommand{\BayesNICERXMMBoth}{0.23 \pm 0.01}

\begin{document} 
\reportnum{LA-UR-21-20534}

\title{Nuclear-Physics Multi-Messenger Astrophysics Constraints on the Neutron-Star Equation of State:\\ Adding NICER's PSR J0740+6620 Measurement}

\correspondingauthor{Peter T.~H.~Pang}
\email{thopang@nikhef.nl}

\author[0000-0001-7041-3239]{Peter T.~H.~Pang}
\affiliation{Nikhef, Science Park 105, 1098 XG Amsterdam, The Netherlands}
\affiliation{Institute for Gravitational and Subatomic Physics (GRASP), Utrecht University, Princetonplein 1, 3584 CC Utrecht, The Netherlands}

\author[0000-0003-2656-6355]{Ingo Tews}
\affiliation{Theoretical Division, Los Alamos National Laboratory, Los Alamos, NM 87545, USA}

\author[0000-0002-8262-2924]{Michael W.~Coughlin}
\affiliation{School of Physics and Astronomy, University of Minnesota,
Minneapolis, Minnesota 55455, USA}

\author[0000-0002-8255-5127]{Mattia Bulla}
\affiliation{The Oskar Klein Centre, Department of Astronomy, Stockholm University, AlbaNova, SE-106 91 Stockholm, Sweden}

\author[0000-0001-6800-4006]{Chris Van Den Broeck}
\affiliation{Nikhef, Science Park 105, 1098 XG Amsterdam, The Netherlands}
\affiliation{Institute for Gravitational and Subatomic Physics (GRASP), Utrecht University, Princetonplein 1, 3584 CC Utrecht, The Netherlands}

\author[0000-0003-2374-307X]{Tim Dietrich}
\affiliation{Institut f\"{u}r Physik und Astronomie, Universit\"{a}t Potsdam, 
Haus 28, Karl-Liebknecht-Str. 24/25, 14476, Potsdam, Germany}
\affiliation{Max Planck Institute for Gravitational Physics (Albert Einstein Institute),
Am M\"{u}hlenberg 1, Potsdam 14476, Germany}

\date{\today}

\begin{abstract}
In the past few years, new observations of neutron stars and neutron-star mergers have provided a wealth of data that allow one to constrain the equation of state of nuclear matter at densities above nuclear saturation density.
However, most observations were based on neutron stars with masses of about 1.4 solar masses, probing densities up to $\sim$ 3-4 times the nuclear saturation density.
Even higher densities are probed inside massive neutron stars such as PSR J0740+6620. 
Very recently, new radio observations provided an update to the mass estimate for PSR J0740+6620 and X-ray observations by the NICER and XMM telescopes constrained its radius. 
Based on these new measurements, we revisit our previous nuclear-physics multi-messenger astrophysics constraints and derive updated constraints on the equation of state describing the neutron-star interior.
By combining astrophysical observations of two radio pulsars, two NICER measurements, the two gravitational-wave detections GW170817 and GW190425, detailed modeling of the kilonova AT2017gfo, as well as the gamma-ray burst GRB170817A, we are able to estimate the radius of a typical 1.4-solar mass neutron star to be $\ROnePointFourNICERXMMBoth$ at 90\% confidence. 
Our analysis allows us to revisit the upper bound on the maximum mass of neutron stars and disfavors the presence of a strong first-order phase transition from nuclear matter to exotic forms of matter, such as quark matter, inside neutron stars.
\end{abstract}



\section{Introduction}
One of the major challenges in modern nuclear physics is the characterisation of the equation of state (EOS) describing matter at supra-nuclear densities. 
These densities are probed inside neutron stars (NSs), which are among the most compact objects in the Universe.
Therefore, NSs are ideal laboratories to test theories of strong interactions at conditions that cannot be realized experimentally on Earth and to validate or falsify theoretical models for the EOS of dense neutron-rich matter (see, e.g.,~\cite{Lattimer:2012nd} and \cite{Ozel:2016oaf}). 

\begin{figure}
    \centering
    \includegraphics[width=1.\columnwidth]{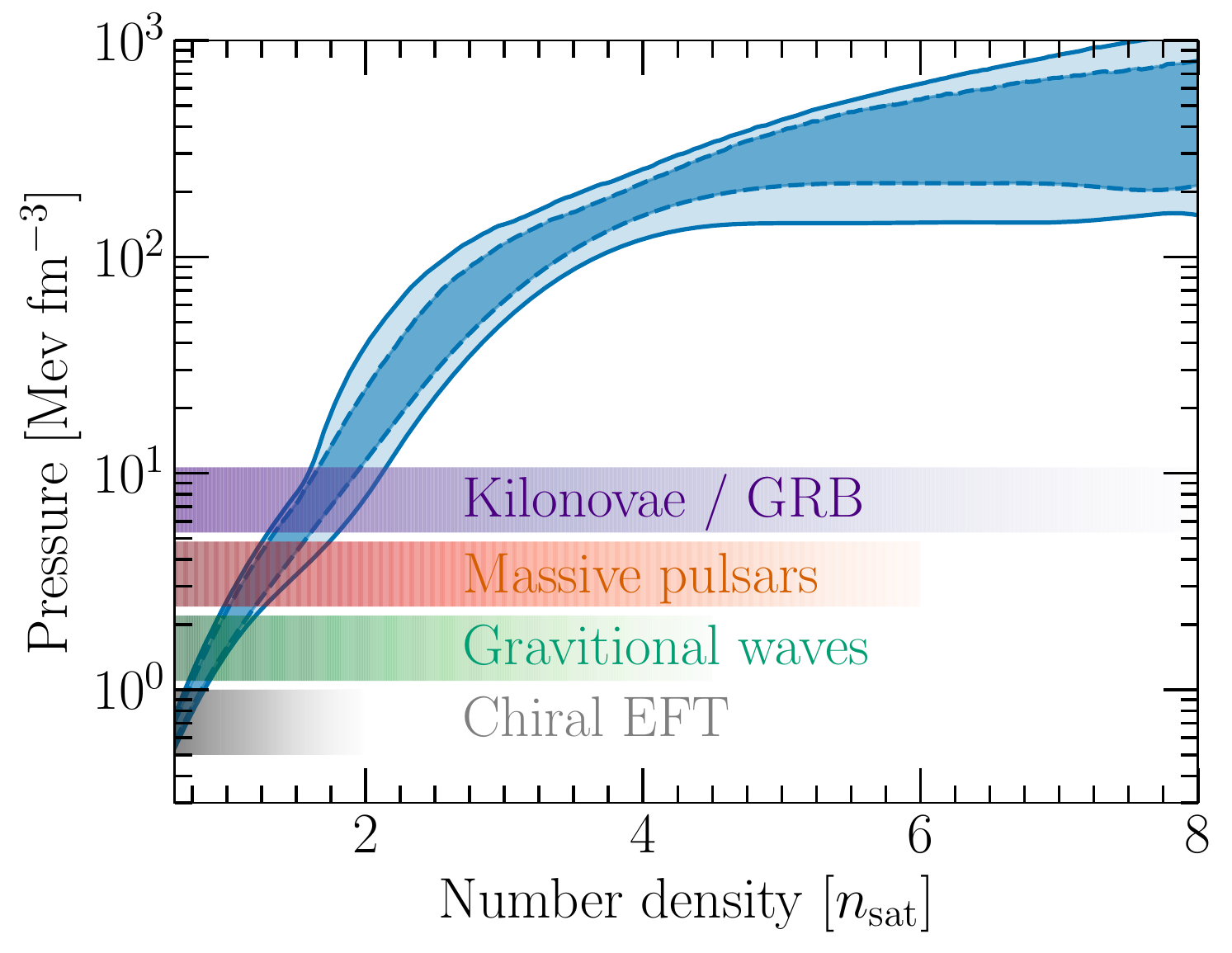}
    \caption{The posterior for the pressure as a function of number density for our final analysis is shown at 68\% and 95\% credible intervals (blue light- and dark-shaded bands, respectively). 
    The shaded bars indicate qualitatively which density regions are probed by different NS information, while the corresponding pressures can be extracted using the pressure vs.\ density band. The bars refer to theoretical modeling at low densities using chiral effective field theory (gray), gravitational waves (green) where the maximum probed density is the central density of GW190425's primary component, massive pulsars (orange) where we give the central density of PSR J0740+6620, and kilonova and GRB (purple).
    Because kilonova and GRB properties depend on the black hole formation mass, and therefore the maximum allowed mass by the EOS, we show the central density of the maximum mass NS.}
    \label{fig:intro}
\end{figure}

In the recent years, a number of new observational constraints on the EOS appeared, either from single NSs, such as the observations of massive pulsars~\citep{Demorest:2010bx, Antoniadis:2013pzd, Arzoumanian:2017puf, Cromartie:2019kug} or the X-ray pulse-profile modeling of J0030+0451 by the Neutron-Star Interior Composition Explorer (NICER)~\citep{Miller:2019cac, Riley:2019yda}, but also from the observation of binary NS mergers via gravitational waves (GWs), GW170817~\citep{TheLIGOScientific:2017qsa,Abbott:2018wiz,Abbott:2018exr,LIGOScientific:2018mvr} and GW190425~\citep{Abbott:2020uma,Abbott:2020niy}, and corresponding electromagnetic (EM) counterparts associated with GW signals, namely AT2017gfo and GRB170817A (e.g.,~\cite{GBM:2017lvd,Arcavi:2017xiz,Coulter:2017wya,Lipunov:2017dwd,Soares-Santos:2017lru,Tanvir:2017pws,Valenti:2017ngx}). 
All these measurements provide key input to analyse the NS structure and the EOS. 
This wealth of data made available by multi-messenger observations of NSs and NS mergers in the last years has energized the field and triggered many exciting studies (e.g.,~\cite{Bauswein:2017vtn, Metzger:2017wot, Radice:2017lry, Ruiz:2017due, Annala:2017llu, Most:2018hfd, Hinderer:2018pei, Radice:2018ozg, Tews:2018chv, Coughlin:2018fis, Essick:2019ldf, Capano:2019eae, Raaijmakers:2019dks, Dietrich:2020lps, Essick:2020flb, Essick:2021kjb}).
However, most of the existing observational data that is sufficiently constraining to improve our understanding of the EOS, e.g., from GW170817 or NICER observations of J0030+0451, probe typical NSs with masses of the order of $1.4 M_{\odot}$. 
Hence, these measurements explore the EOS `only' up to densities of the order of three to four times the nuclear saturation density, $n_{\rm sat}\sim 0.16\,\rm{fm}^{-3}$; cf.~Fig.~\ref{fig:intro}. 
These `intermediate' densities are likely below the onset of a possible phase transition to exotic forms of matter, e.g., quark matter (see, e.g.,~\citealt{Annala:2019puf}).

To explore the high-density EOS and to place constraints on the possible existence of a phase transition, it is crucial to observe isolated NSs close to the maximum mass supported by the EOS. 
Alternatively, the high-density part of the EOS can also be probed through binary NS mergers.
Once the two stars merge, they can potentially form a hypermassive remnant even exceeding the maximum mass of individual NSs~\citep{Radice:2016rys,Margalit:2017dij, Ruiz:2017due,Rezzolla:2017aly}. 
However, such remnants have not yet been observed through GWs but only through the associated EM observations which, due to their more involved interpretation, leads to larger uncertainties for the EOS (e.g., \cite{Heinzel:2020qlt} and \cite{Kawaguchi:2019}).
Fortunately, the recently published second observation by NICER \citep{Wolff:2021oba,Miller:2021qha,Riley:2021pdl} provides a crucial new data point for an isolated NS close to the maximum mass. 

NICER is a NASA mission on board the International Space Station that measures the X-ray pulse profile of selected pulsars which allows one to extract information on the configuration of X-ray hot spots.
Additionally, the pulse profile is sensitive to the light bending around the pulsar (see Sec. 4 in ~\citealt{Wolff:2021oba}), and therefore, provides information on the NS compactness, which in turn allows to extract data on the NS mass and radius. 
The first NICER measurement was reported in December 2019 and targeted the pulsar J0030+0451, for which both mass and radius were unknown.
Two independent analyses of the first NICER observation provided mass-radius constraints for this NS of $1.34^{+0.15}_{-0.16} M_{\odot}$ and $12.71^{+1.14}_{-1.19}$~km~\citep{Riley:2019yda} or $1.44^{+0.15}_{-0.14} M_{\odot}$ and $13.02^{+1.24}_{-1.06}$~km~\citep{Miller:2019cac} at 68\% confidence.

In its second observation, NICER analyzed X-ray data from the millisecond pulsar PSR J0740+6620 \citep{Miller:2021qha,Riley:2021pdl}. 
This NS is the heaviest NS observed to date with a known mass of $2.08\pm 0.07\, M_{\odot}$~(\citealt{Fonseca:2021wxt}, 
updated from its original value reported by \citealt{Cromartie:2019kug}).
Combining the known mass with the pulse-profile modeling allowed the NICER collaboration to measure the radius of PSR J0740+6620.
Two independent analyses by the NICER collaboration found the radius to be $12.39^{+1.30}_{-0.98} \rm km$ \citep{Riley:2021pdl} or $13.71^{+2.61}_{-1.50}\rm km$ \citep{Miller:2021qha} at 68\% confidence. 
While the NICER data provide information about the modulated emission from the star, the analyses of \cite{Miller:2021qha} and \cite{Riley:2021pdl} additionally used information from the X-ray Multi-Mirror (XMM)-Newton telescope \citep{Struder:2001bh, Turner:2000jy} to improve the total flux measurement from the star, due to a smaller rate of background counts.

\begin{figure}
    \centering
    \includegraphics[width=1.\columnwidth]{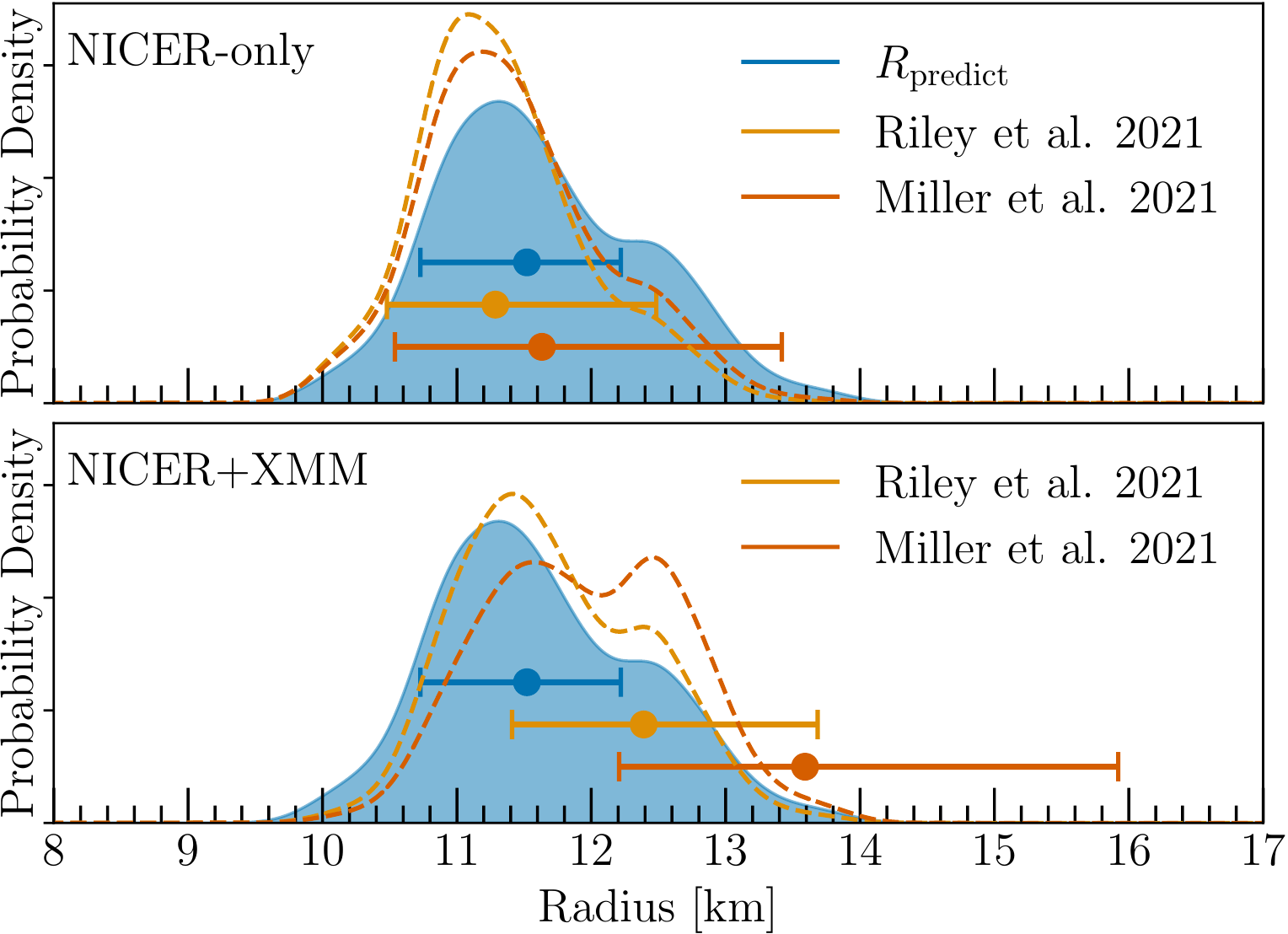}
    \caption{Upper panel: Posterior distribution function for $R_{\rm predict}$ from the NMMA framework of \cite{Dietrich:2020lps} (blue shaded). 
    The median and 68\% uncertainty for the radius prediction of PSR J0740+6620 are shown as blue error bar. 
    We also show the NICER-only measurement of \cite{Riley:2021pdl}(yellow) and \cite{Miller:2021qha}(red) at 68\% uncertainty. 
    The posteriors after the inclusion of the updated observations of PSR J0740+6620 are shown by dashed lines.
    Lower panel: Similar to the upper panel but including also XMM data.}
    \label{fig:predict_compare}
\end{figure}

In this article, we incorporate the new X-ray observation of PSR J0740+6620 and its updated mass within our existing nuclear-physics multi-messenger astrophysics (NMMA) framework, which we have developed and described in \cite{Dietrich:2020lps}.
This allows us to revisit our constraints on the NS EOS, in particular on the existence of strong first-order phase transitions, the maximum mass of neutron stars, and the nature of GW190814~\citep{Tews:2020ylw}.

This paper is structured as follows. 
In Section~\ref{sec:previous} we briefly summarize our previous results and review recent works including the NICER measurement of PSR J0740+6620.
In Section~\ref{sec:meth}, we review our NMMA framework.
Using the new NICER data, in Section~\ref{sec:results}, we discuss our prediction for the NS EOS (Sec.~\ref{sec:EOS}), the NS maximum mass and the probability for GW190814~\citep{Abbott:2020khf} to be a black hole-NS merger (Sec.~\ref{sec:Mmax}), and investigate to what extent the recent NICER observations informs us about the existence of a phase transition in the EOS (Sec.~\ref{sec:PT}). 
We will give a summary of our results in Section~\ref{sec:summary}. 

\section{Previous works}~\label{sec:previous}
In our previous work~\citep{Dietrich:2020lps}, we included the pulsar mass measurements of PSR J0740+6620, PSR J0348+4042, and PSR J1614-2230~\citep{Cromartie:2019kug, Antoniadis:2013pzd, Arzoumanian:2017puf, Demorest:2010bx}, 
GW data from the NS mergers GW170817 and GW 190425, 
information from the kilonova AT2017gfo, the gamma-ray burst GRB170817A as well as its afterglow~\citep{GBM:2017lvd}, 
and the NICER observation of PSR J0030+0451~\citep{Riley:2019yda, Miller:2019cac} in a Bayesian Inference framework based on systematic nuclear-physics input from chiral effective field theory (EFT).
We obtained a radius of a typical NS of $R_{1.4}=11.75^{+0.55}_{-0.50}$~km~\citep{Dietrich:2020lps} at 68\% uncertainty.
Based on these results, our prediction of the radius of PSR J0740+6620 was $R_{\rm predict} = 11.52^{+0.70}_{-0.79}\rm km$ at 68\% confidence level. 
We compare this prediction with the recent NICER measurements in Fig.~\ref{fig:predict_compare}.
We find that our estimate is in excellent agreement with the results obtained in \cite{Riley:2021pdl} and \cite{Miller:2021qha} using only the NICER data.
Once data from the XMM-Newton observatory is additionally taken into account, the radius is pulled to larger values decreasing the agreement between our prediction and the measurement, but deviations are $\lesssim 1\sigma$.
This effect is stronger for the Maryland-Illinois result \citep{Miller:2021qha}, which appears to be caused by a number of differences between the individual analyses of \cite{Miller:2021qha} and \cite{Riley:2021pdl}: 
differences in the prior on the cross-calibration uncertainty of NICER and XMM-Newton which is allowed to be two times larger than the measured deviation in \cite{Miller:2021qha} but an order of magnitude larger in the analysis of \cite{Riley:2021pdl}, 
differences in the radius prior which has an upper bound of 16~km in the analysis of \cite{Riley:2021pdl} and approximately 25~km in \cite{Miller:2021qha},
differences in the sampling algorithms and their convergence that affected the posterior widths,
and differences in the assumed distribution of the blank-sky counts to estimate the XMM background.

Since the first announcement of the NICER results for PSR J0740+6620~\citep{Nicer_pressrelease} there have been several studies of the implications of this radius measurement~\citep{Somasundaram:2021ljr, Biswas:2021yge, Li:2021thg, Annala:2021gom}. 
However, these studies did not use the full posterior samples released in~\cite{Riley:2021pdl,Miller:2021qha}, but employed hard cuts for the radius-mass measurement, which can lead to biases during the final multi-messenger analysis, as shown in, e.g.,~\cite{Miller:2019nzo}. 
In more detail, the studies in \cite{Somasundaram:2021ljr} and \cite{Biswas:2021yge} were based on phenomenological nuclear-physics descriptions, but did not include systematic nuclear-theory calculations with uncertainty estimates, e.g., from chiral EFT, to constrain the low-density EOS.
Instead, \cite{Somasundaram:2021ljr} studied the impact of the NICER observation of PSR J0740+6620 on the existence of a first-order phase transition or quarkyonic matter using two different models for the high-density part of the EOS.
They found that the NICER observation of PSR J0740+6620 cannot rule out first-order phase transitions, but this study also did not systematically include other astrophysical constraints in a Bayesian analysis.
Using a different EOS parametrization, \cite{Biswas:2021yge} combined a hypothetical radius measurement of PSR J0740+6620 with previous GW and NICER observations and the recent result for the neutron-skin thickness of $^{208}$Pb from the PREX-II experiment~\citep{Adhikari:2021phr}.
In contrast to these works, \cite{Annala:2021gom} used theoretical nuclear-physics input from chiral EFT at low densities and perturbative Quantum Chromodynamics (QCD) at high densities to constrain the EOS.
However, they did not perform a Bayesian analysis to constrain the EOS given available astrophysical data, but instead implemented various constraints using hard cuts; see above.
None of these papers included multi-messenger constraints from a detailed modeling and parameter estimation of EM counterparts associated with binary NS mergers, e.g., from a Bayesian Inference of AT2017gfo and GRB170817A.

\cite{Raaijmakers:2021uju} and \cite{Miller:2021qha} were directly based on the NICER and XMM measurements of PSR J0740+6620 and studied the influence of the new NICER data using the full posterior samples.
For this purpose, \cite{Miller:2021qha} employed very general and conservative EOS priors that were not directly informed by nuclear-theory calculations at low densities.
In addition to the NICER observations of pulsars PSR J0030+0451 and PSR J0740+6620, final EOS constraints in \cite{Miller:2021qha} used other heavy-pulsar mass measurements, gravitational-wave observations of GW170817 and GW190425, and constraints on the nuclear symmetry energy.
Results presented in \cite{Raaijmakers:2021uju} were instead constrained by chiral EFT calculations up to $1.1 n_{\rm sat}$, comparing four different chiral EFT calculations. 
In addition, EM information from AT2017gfo were included as well as information from GW170817 and GW190425. 

In this paper, in contrast to the studies presented in \cite{Miller:2021qha} and \cite{Raaijmakers:2021uju}, we use updated GW models~\citep{Dietrich:2019kaq} and different kilonova models with detailed microphysical input that also explore deviations from spherical symmetry~\citep{Kasen:2017sxr,Bulla:2019muo}; cf.\ \cite{Heinzel:2020qlt} and \cite{Dietrich:2020lps} for details about systematic uncertainties of kilonova modelling.
In case of \cite{Miller:2021qha}, we also include low-density input from chiral EFT.


\section{Methodology}~\label{sec:meth}

Our NMMA framework uses Bayesian inference tools to analyze a set of EOSs with respect to their agreement with several astrophysical observations of NSs.

The initial EOS set is constrained at low densities by calculations of the energy per particle of neutron matter using interactions from chiral EFT~\citep{Epelbaum:2008ga, Machleidt:2011zz}.
Chiral EFT is a low-energy systematic theory for nuclear forces and provides a momentum expansion of two-nucleon and multi-nucleon interactions. 
Interactions are arranged in an order-by-order scheme that organizes various interaction mechanisms according to their relative importance.
By going to higher orders, the precision of the calculation is improved at the cost of more involved calculations. 
An important benefit of the chiral EFT scheme is that it allows one to estimate theoretical uncertainties~\citep{Epelbaum:2015epja, Drischler:2020yad}.
Given chiral EFT Hamiltonians describing the strong interactions between nucleons, a many-body method is required to solve the many-body Schr\"{o}edinger equation and calculate the energy per particle.
The EOS set used here is constrained by calculations employing Quantum Monte Carlo (QMC) methods which are among the most precise methods to solve the nuclear many-body problem~\citep{Carlson:2015} and have been combined with chiral EFT interactions with great success~(e.g., \cite{Gezerlis:2013, Piarulli:2017dwd, Lonardoni:2017hgs, Lynn:2019rdt}).
Specifically, here we use the QMC calculations by \cite{Tews:2018kmu} using local chiral EFT interactions from \cite{Gezerlis:2013, Gezerlis:2014, Tews:2015ufa}, and \cite{Lynn:2016}, which agree very well with other microscopic calculations based on chiral EFT~\citep{Huth:2020ozf}. 
Furthermore, \cite{Raaijmakers:2021uju} have shown that EOS constraints are only weakly dependent on the choice of a particular chiral EFT calculation at low densities.

The range of applicability of chiral EFT calculations is limited because the average nucleon momentum increases with density and the momentum expansion breaks down.
While the exact breakdown density for individual chiral EFT interactions is unknown, it was estimated to be between $1-2 n_{\rm sat}$ for the QMC calculations used here~\citep{Tews:2018kmu, Essick:2020flb}.
We constrain the NS EOS with QMC calculations of the EOS up to a density of $1.5 n_{\rm sat}$ but results were found to show only a very weak dependence to a variation of this density between $1-2 n_{\rm sat}$ \citep{Essick:2020flb}.
We use the speed-of-sound extension scheme from \cite{Tews:2018chv} to extend the EOS to higher densities in a general way. 
The speed-of-sound extension scheme allows us to explore the physically plausible EOS space without making any strong model assumptions (see also \cite{Greif:2018njt} and \cite{Raaijmakers:2021uju}). 
We only require the EOSs to explore speeds of sound, $c_S$, limited by $0\leq c_S \leq 1$ in units of the speed of light, and to provide a maximum mass for NSs of at least 1.9 $M_{\odot}$.
Our EOS set explicitly includes EOS with regions of sudden stiffening or softening, e.g., strong first-order phase transitions towards exotic forms of matter.
For our EOS set, we impose a uniform prior on the radius of a typical 1.4$M_{\odot}$ NS, $R_{1.4}$. 
To estimate the impact of the particular choice of the EOS prior, we have also investigated an EOS prior without this additional requirement, see Tab.~\ref{tab:results_natural} of the Appendix

As a next step, we analyze our EOS set with respect to available NS observations. 
We start by incorporating a constraint on the maximum mass of NSs through the radio observations of the heaviest pulsars known to date,  PSR~J0348+4042~\citep{Antoniadis:2013pzd} and PSR~J1614-2230~\citep{Arzoumanian:2017puf}. 
The existence of these pulsars can only be explained if the NS EOS supports masses that lie above the individual masses of these pulsars.
Hence, radio pulsar measurements of heavy NSs provide a lower bound on the maximum NS mass and the constraints of the high-density EOS;
cf.~Fig.~\ref{fig:intro}.
We stress that at this stage we do not include the mass measurement of PSR J0740+6620, because information will be included through the new NICER measurement.
An upper bound of the maximum NS mass follows from the EM observation of GRB170817A and AT2017gfo. 
As outlined in, e.g., \cite{Margalit:2017dij}, the observed EM signatures indicate the formation of a black hole as the final product of the binary NS merger GW170817. 
Combining this information with the estimated total remnant mass from the GW observation leads to a non-rotating maximum NS mass of $M_{\rm max} \leq 2.16^{+0.17}_{-0.15} M_\odot$~\citep{Rezzolla:2017aly}.
Incorporating the constraints on the maximum mass leads to a re-weighting of the original chiral EFT EOS set.

As a next step, we include the NICER measurement of PSR J0030+0451, for which the inferred mass-radius posterior probability distributions was not dominated by systematic uncertainties and inferred system parameters were in agreement for different analyses~\citep{Miller:2019cac, Riley:2019yda}.
Finally, we use the resulting EOS set for GW and kilonova parameter estimation following the methods outlined in~\cite{Dietrich:2020lps}. 

\begin{figure*}[t]
    \centering
    \includegraphics[width=0.46\textwidth]{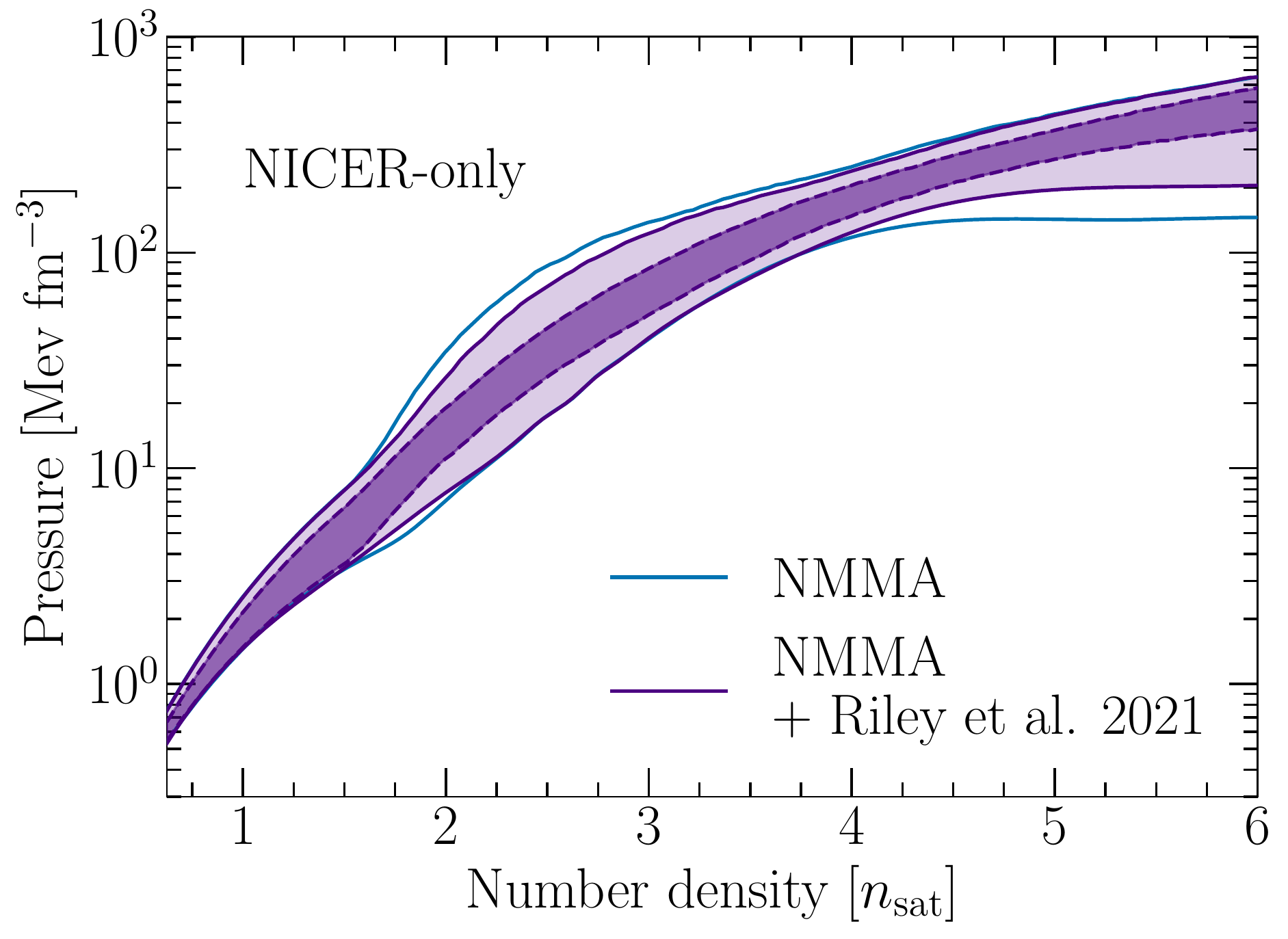}
    \includegraphics[width=0.5\textwidth]{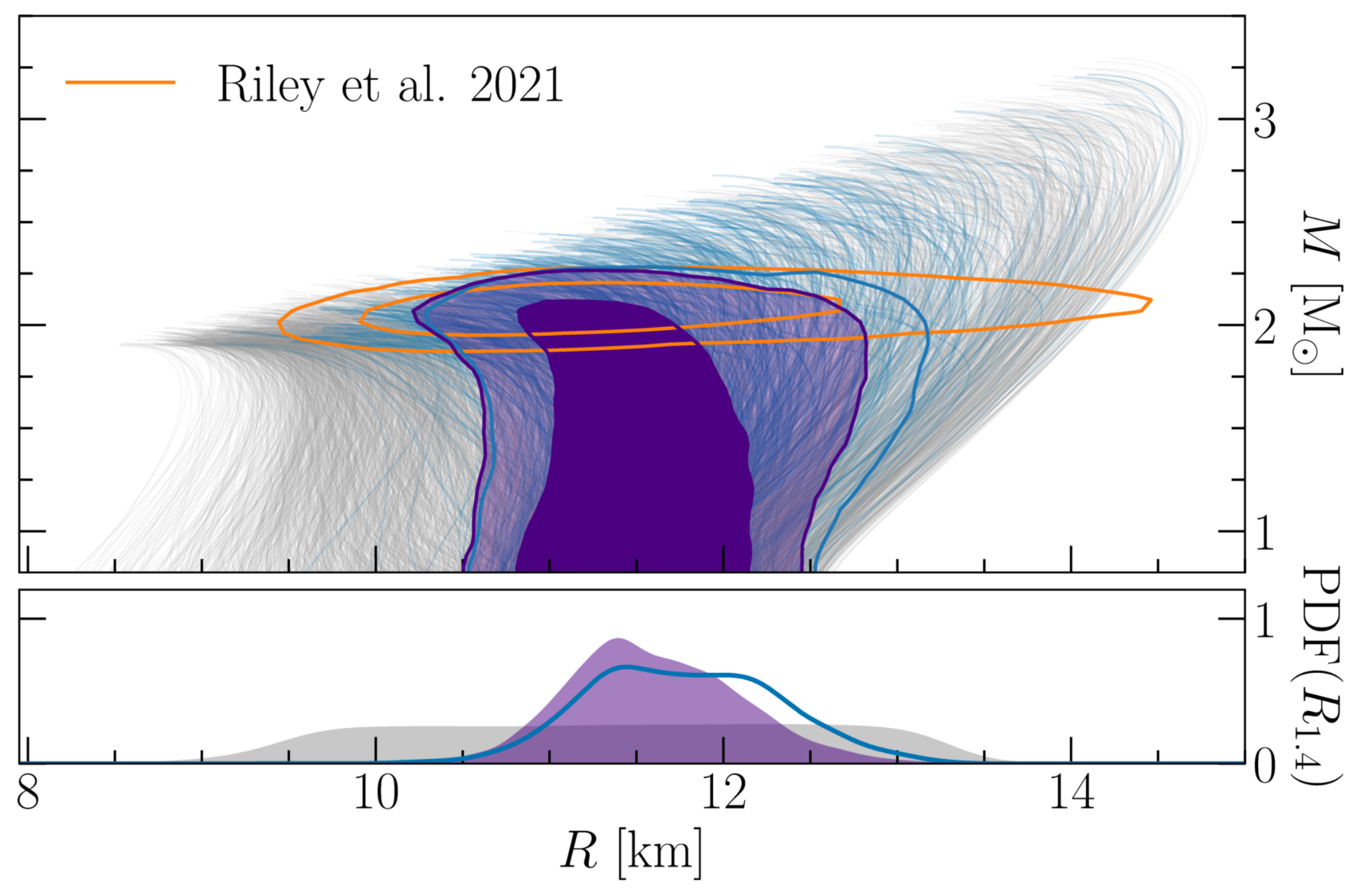}\\
    \includegraphics[width=0.46\textwidth]{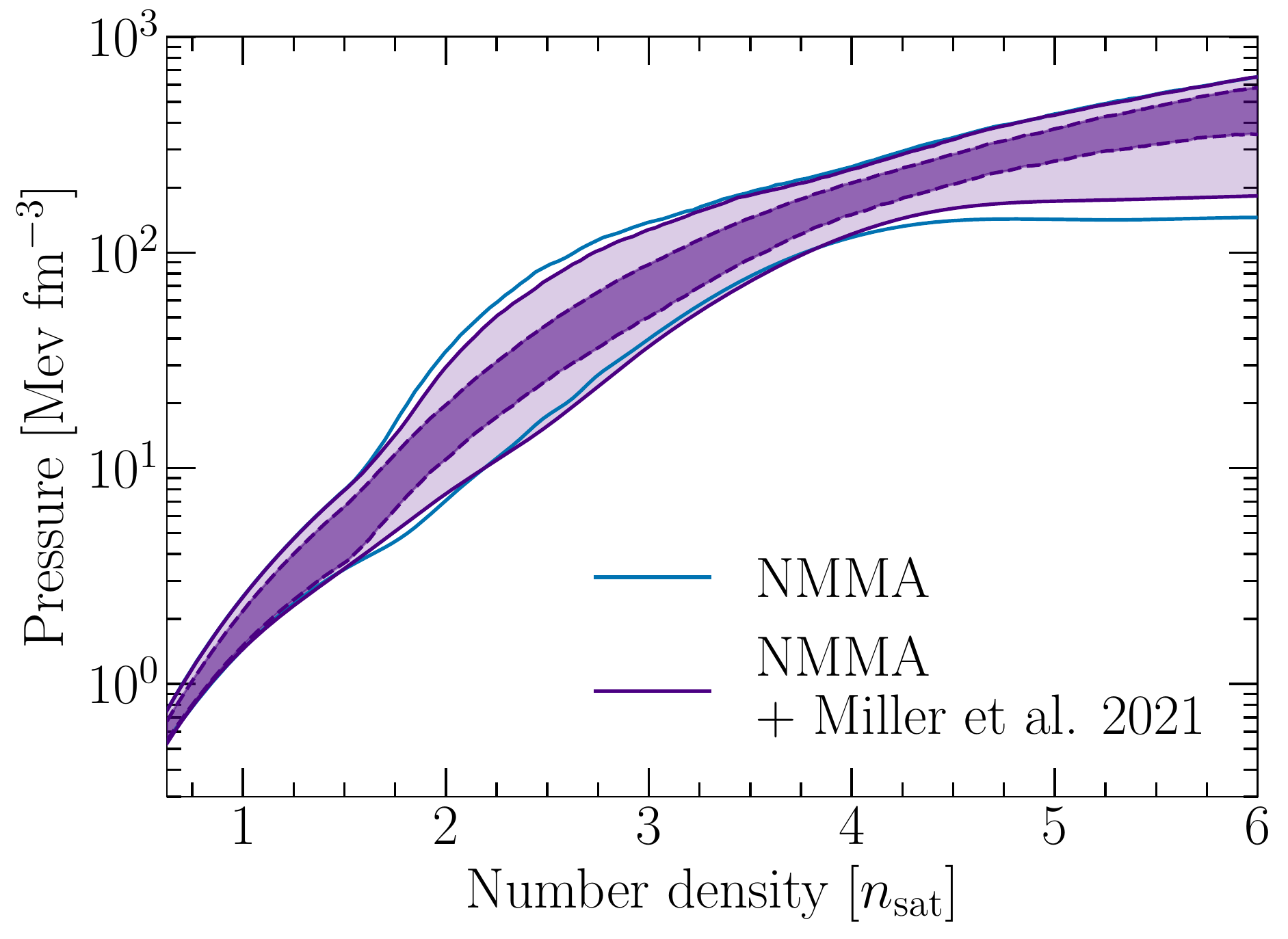}
    \includegraphics[width=0.5\textwidth]{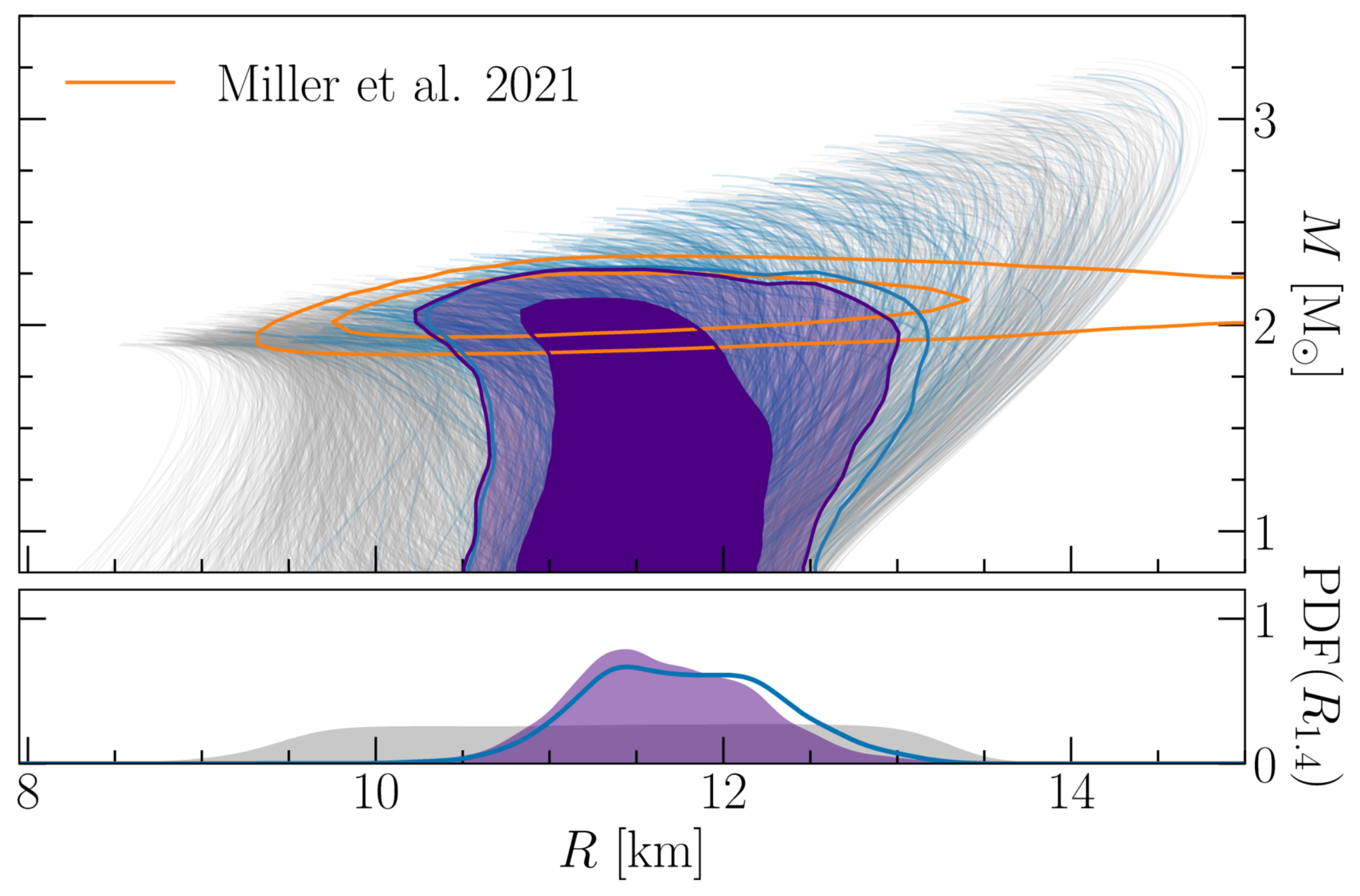}
    \caption{Left panels: The posterior for the pressure as a function of number density including the NICER-only observation of PSR J0740+6620 from \cite{Riley:2021pdl} (upper panel) and from \cite{Miller:2021qha} (lower panel). 
    The bands indicate 68\% and 95\% credible intervals. 
    The 95\% band for the NMMA result without the new NICER measurement is shown as comparison (blue line).
    Right panels: NICER mass-radius posteriors of PSR J0740+6620 plotted at 68\% and 95\% confidence intervals (orange contours) and the EOSs included in the analysis (gray lines). 
    The 95\% contour for the NMMA result without including the new NICER observation is shown as thick blue line, while the individual EOSs within this credible interval of the NMMA analysis are shown as thin blue lines. 
    The resulting mass-radius posterior after the inclusion of the new NICER-only observation is shown in purple for the NICER result of \cite{Riley:2021pdl} (upper panel) and of \cite{Miller:2021qha} (lower panel)at  68\% and 95\% credible intervals.
    The 1D insets show the posteriors for $R_{1.4}$ with (purple) and without (blue) the inclusion of the NICER-only measurement of PSR J0740+6620.}
    \label{fig:MR_R14_Mmax_PDF}
\end{figure*}

\begin{figure*}[t]
    \centering
    \includegraphics[width=0.46\textwidth]{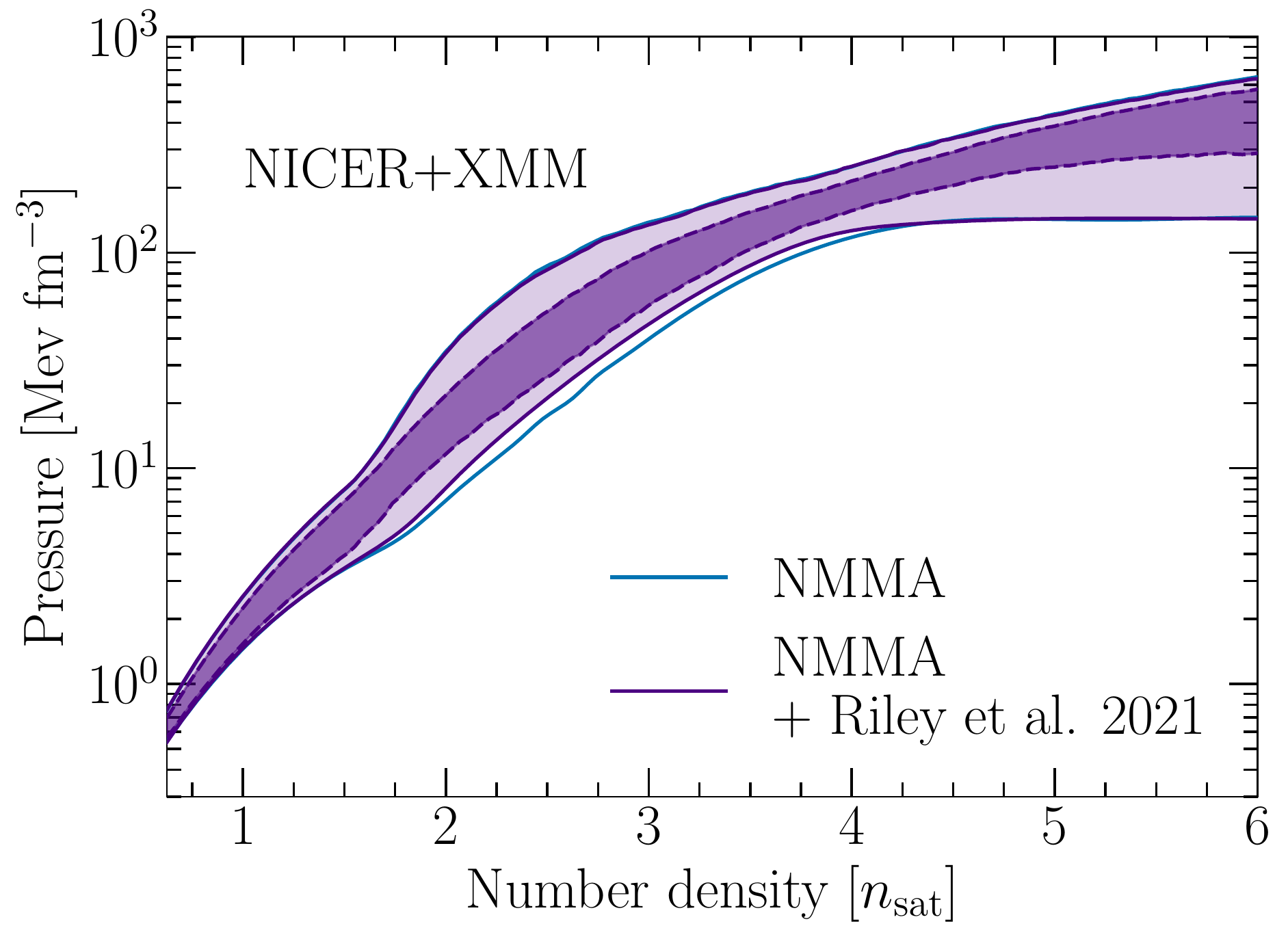}
    \includegraphics[width=0.5\textwidth]{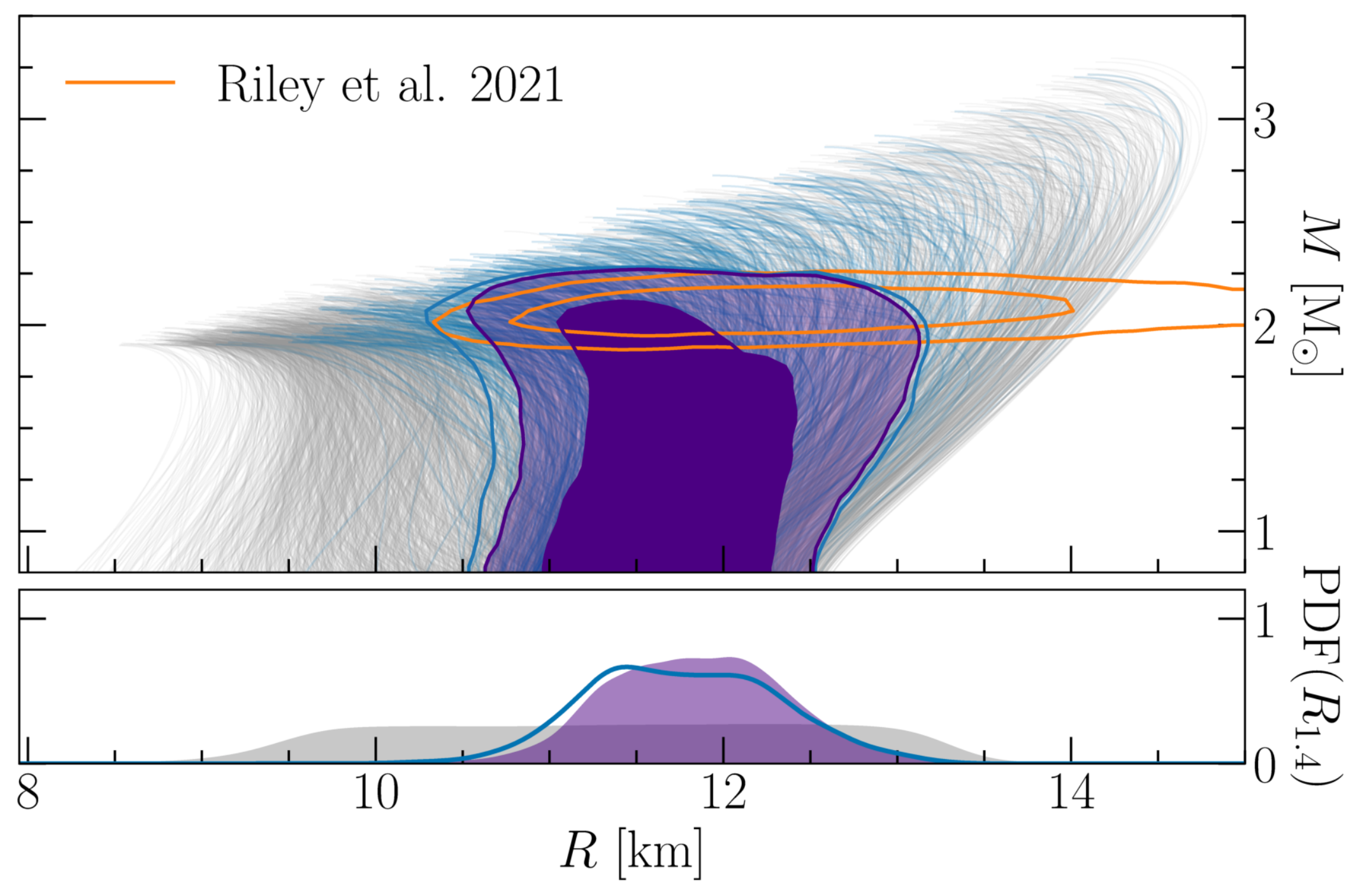}\\
    \includegraphics[width=0.46\textwidth]{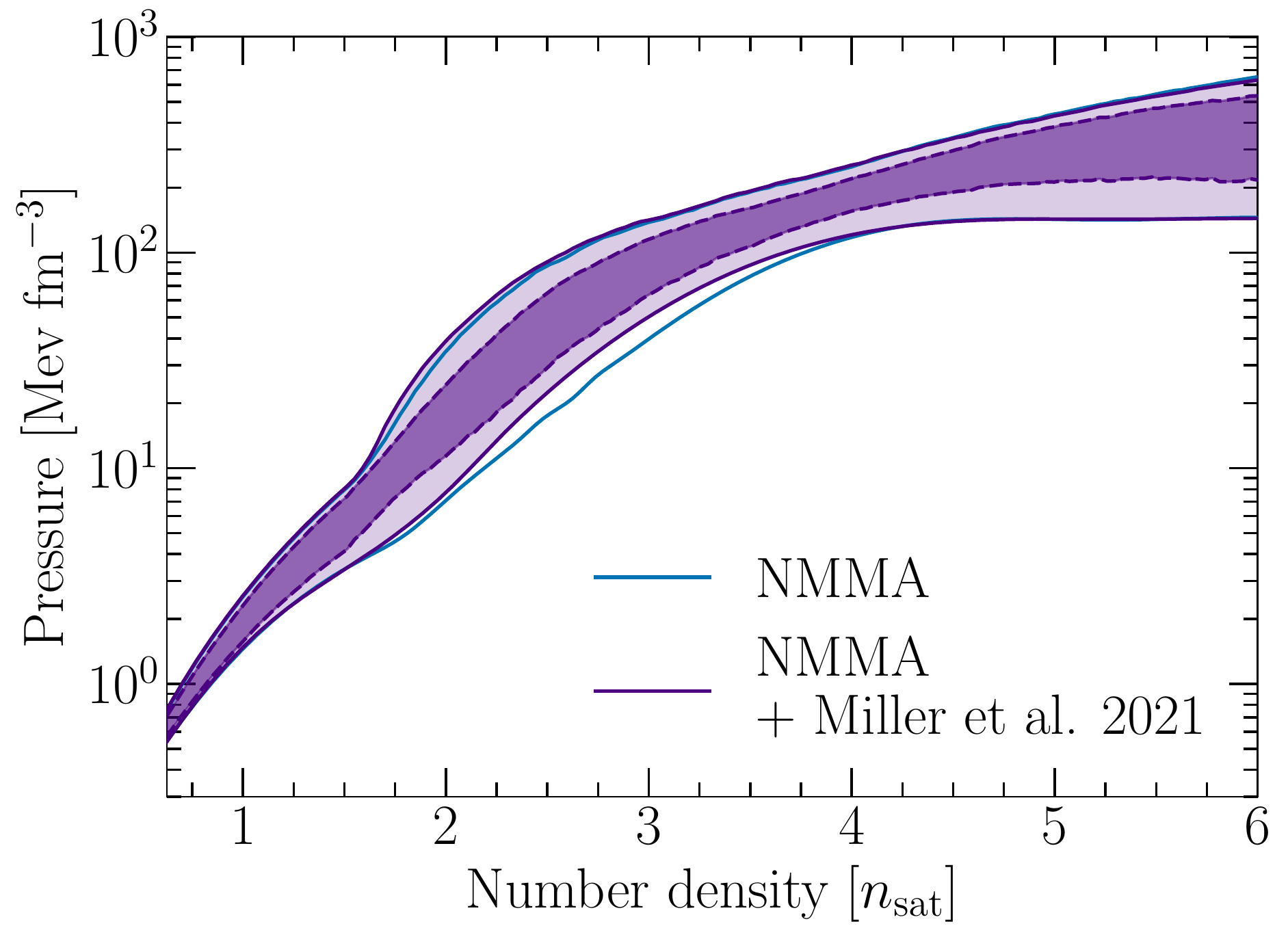}
    \includegraphics[width=0.5\textwidth]{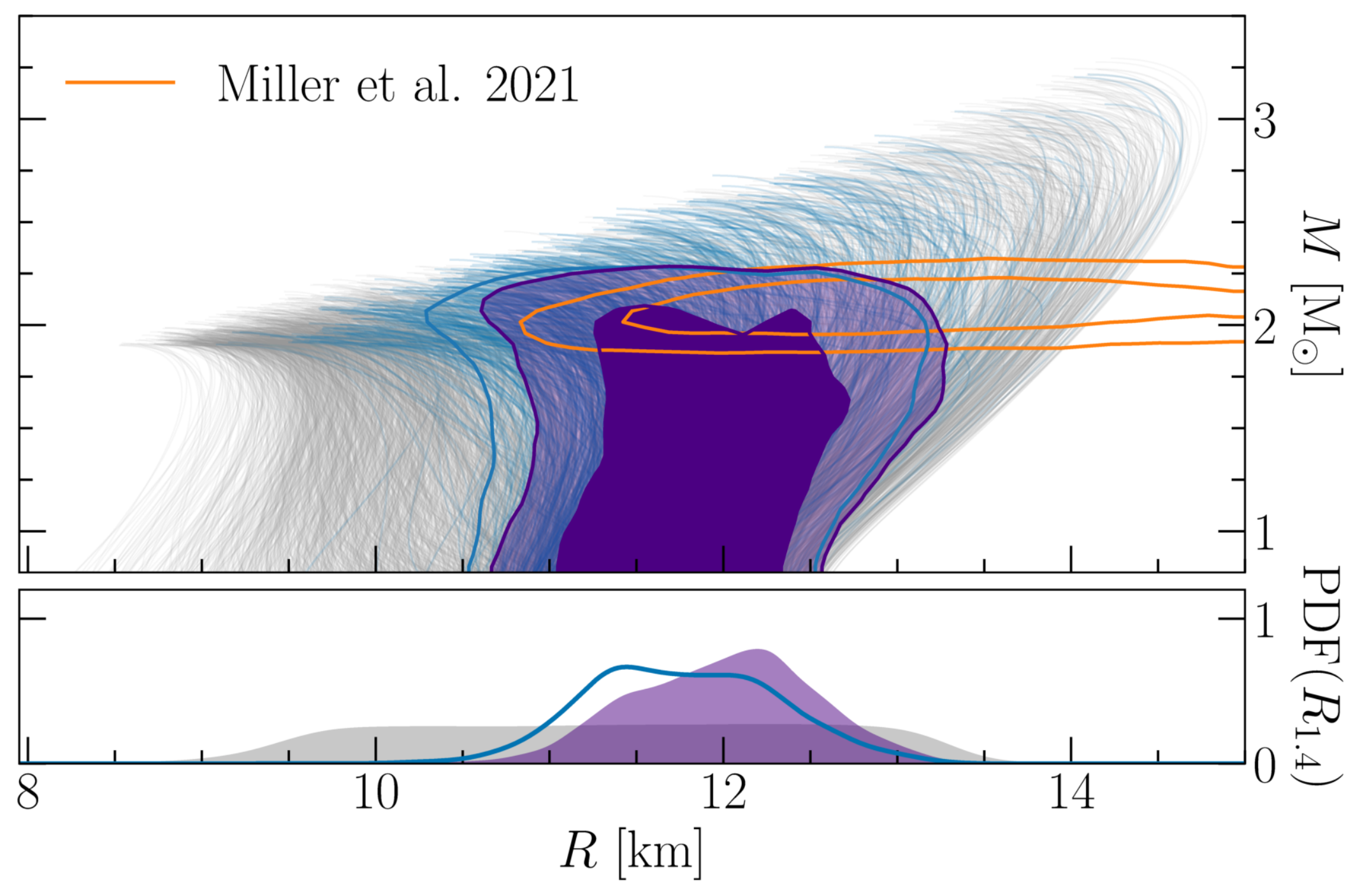}
    \caption{Same as Fig.~\ref{fig:MR_R14_Mmax_PDF} but using the NICER+XMM data.}
    \label{fig:MR_R14_Mmax_PDF_XMM}
\end{figure*}

We now include the NICER observations of PSR~J0740+6620 \citep{Miller:2021qha, Riley:2021pdl}, which are based on a Bayesian inference approach to analyze the energy-dependent thermal X-ray signal of PSR~J0740+6620.
We employ the posterior samples obtained with the two-circular, uniform-temperature spot model from~\cite{miller_m_c_2021_4670689, Miller:2021qha} and the two disjoint , uniform-temperature spots model from~\cite{Riley:2021pdl, riley_thomas_e_2021_4697625}. 
These models provide best agreement with the observed NICER and NICER+XMM data and, for the latter, constrain the radius of PSR~J0740+6620 with a mass of $2.08\pm 0.07\, M_{\odot}$~\citep{Cromartie:2019kug,Fonseca:2021wxt} to be $13.71^{+2.61}_{-1.50}{\rm km}$ and $12.39^{+1.30}_{-0.98}{\rm km}$ at 68\% confidence for \cite{Miller:2021qha} and \cite{Riley:2021pdl}, respectively.

The corresponding likelihood $\mathcal{L}_{\textrm{NICER}}$ is given by
\begin{equation}
\begin{aligned}
	\mathcal{L}_{\textrm{NICER}}(\textrm{EOS}) &= \int d\!M d\!R\ p_{\textrm{NICER}}(M, R)\pi(M, R |\textrm{EOS})\\
	&\propto \int d\!M d\!R\ p_{\textrm{NICER}}(M, R)\delta(R-R(M,\textrm{EOS}))\\
	&\propto \int d\!M\ p_{\textrm{NICER}}(M, R=R(M, \textrm{EOS})),
\end{aligned}
\end{equation}
where $p_{\textrm{NICER}}(M, R)$ is the joint-posterior probability distribution of mass and radius of PSR J0740+6620 as measured by NICER and we use the fact that the radius is a function of mass for a given EOS.\\


\section{Results}~\label{sec:results}

In the following, we discuss the results of our NMMA framework when the new NICER measurement is included.
We give results using constraints from the X-PSI analysis by the Amsterdam group~\citep{Riley:2021pdl} or using the analysis of the Illiniois-Maryland group~\citep{Miller:2021qha} outside/inside of parentheses. The combined results refer to an analysis using the average of the two $(M,R)$ posterior distributions for PSR J0740+6620.
Our findings are summarized in Tab.~\ref{tab:results}.

\begin{figure*}
    \centering
    \includegraphics[width=1\columnwidth]{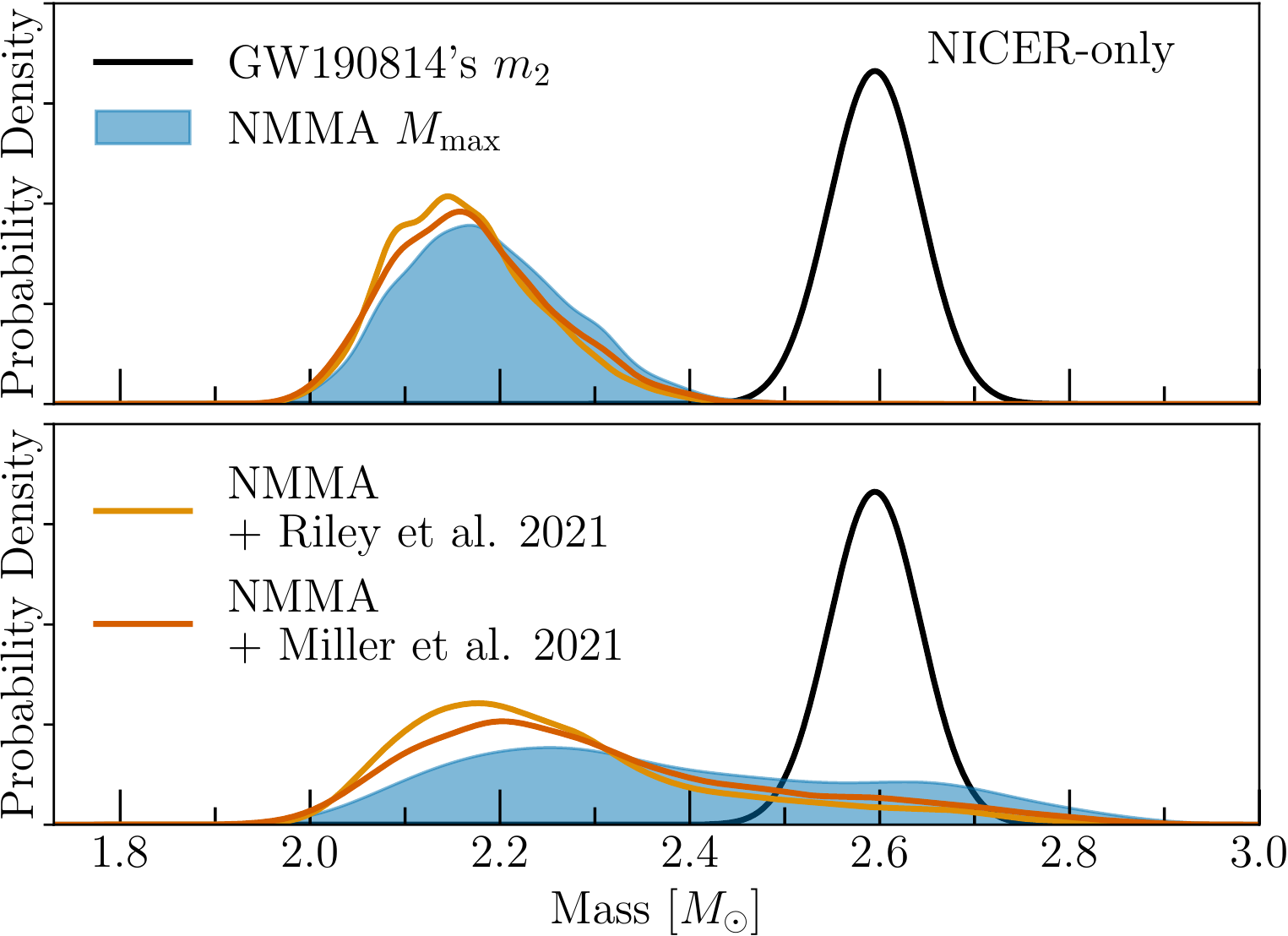}
    \includegraphics[width=0.983\columnwidth]{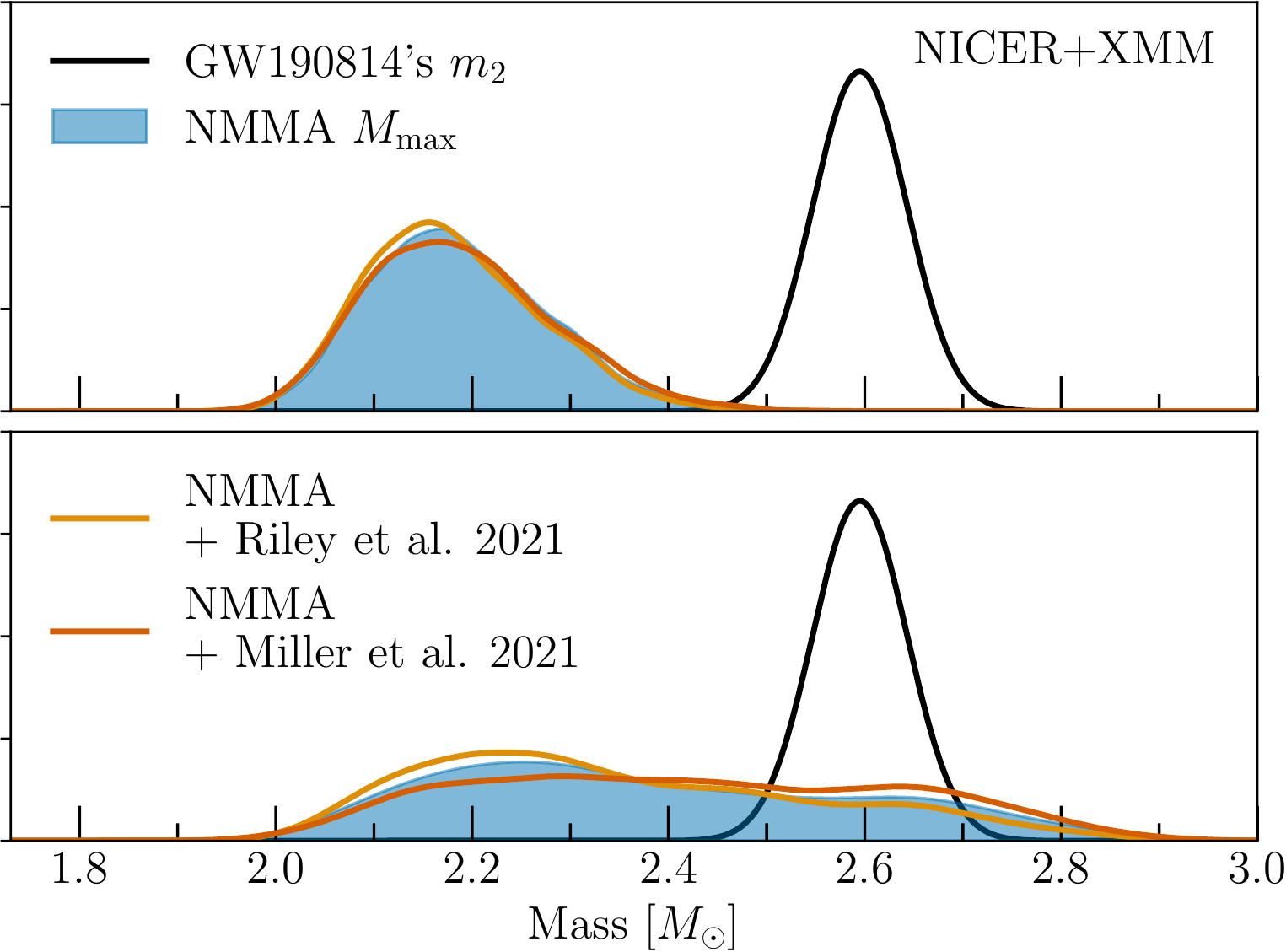}
    \caption{Distributions of $M_{\rm{max}}$ without the new NICER observation (blue bands) and when including the posterior from \cite{Riley:2021pdl} (yellow lines) and \cite{Miller:2021qha} (red lines).
    We show results for the NICER-only data (left panels) and for the NICER+XMM data (right panels), and including the $M_{\rm{max}}$ upper limit suggested in~\cite{Rezzolla:2017aly} (upper panels) and when this upper limit on $M_{\rm{max}}$ is relaxed (lower panels). 
    }
    \label{fig:Mmax_PDF}
\end{figure*}

\subsection{Neutron-star equation of state}~\label{sec:EOS}

In Fig.~\ref{fig:MR_R14_Mmax_PDF}, we show the EOS and mass-radius posteriors after the inclusion of the radius measurement of PSR J0740+6620 using only NICER data.
In this case, the NICER radius measurement shows excellent agreement with the NMMA prediction for the radius of PSR J0740+6620, see also Fig.~\ref{fig:predict_compare}. 
Because the NICER-only data slightly prefers softer EOS in the NMMA set, we observe a softening of our total EOS posterior. 
This can also be seen from the posteriors for the mass-radius relation, that is slightly shifted to lower radii. 
For example, the NMMA framework predicts the radius of a $1.4 M_{\odot}$ NS, $R_{1.4}$, to be $11.75^{+0.86}_{-0.81}\rm km$ without the new NICER measurement. 
Including the measurement, we find $\ROnePointFourNICERRiley\ (\ROnePointFourNICERMiller)$ and a combined result of $\ROnePointFourNICERBoth$ at $90\%$ credibility. 
The median predictions change minimally, by $\sim 200$m, and the uncertainties improve slightly from 4.5\% to 4.2\% for the combined result at 68\% credible interval (and from 7.1\% to 6.9\% at 90\% credible interval). 
Similarly, the radius posteriors of PSR J0740+6620 after including the NICER-only data are shown in Fig.~\ref{fig:predict_compare}. 
The estimated radius changes from $11.52^{+0.70}_{-0.79}{\rm km}$ to $11.26^{+0.56}_{-0.63}{\rm km}$ ($11.35^{+0.61}_{-0.72}{\rm km}$) at 68\% credibility. 

The situation is different when the XMM data is added.
In Fig.~\ref{fig:MR_R14_Mmax_PDF_XMM}, we show our results for the EOS and the mass-radius relation when including the NICER and XMM data. 
Now, these measurements predict larger radii compared to our initial estimation for PSR J0740+6620, see Fig.~\ref{fig:predict_compare}.
This slightly shifts our EOS posterior to the stiffer end and the mass-radius relation to larger radii.  
By including the NICER-XMM measurement, $R_{1.4}$ changes from $11.75^{+0.86}_{-0.81}\rm km$ without the new data to $\ROnePointFourNICERXMMRiley\ (\ROnePointFourNICERXMMMiller)$ and a combined result of $\ROnePointFourNICERXMMBoth$. 
Though the measured radii are well above the NMMA prediction, the uncertainties are sizable and, hence, the radius measurement of PSR J0740+6620 does not change our EOS results significantly. 
The medians shift to slightly larger values but remain statistically consistent with a comparable uncertainty.
Similarly, the NMMA radius prediction for PSR~J0740+6620 changes from $11.52^{+0.70}_{-0.79}{\rm km}$ without the new data to $11.63^{+0.66}_{-0.79}{\rm km}$ ($11.96^{+0.80}_{-0.75}{\rm km}$) at 68\% credibility.

\subsection{Neutron-star maximum mass and GW190814}~\label{sec:Mmax}

The radius measurement of PSR J0740+6620, and its impact on the EOS, allows us to revisit our estimate for the maximum allowed mass for NSs, $M_{\rm max}$, which we found to be $M_{\rm max}=2.18^{+0.14}_{-0.13}M_{\odot}$ without including the new NICER data~\citep{Tews:2020ylw}, see Fig.~\ref{fig:Mmax_PDF}.
Please note that this original estimate used the previously larger mass for PSR J0740+6620 from \cite{Cromartie:2019kug}, which was updated to a lower value in \cite{Fonseca:2021wxt}.

Because the NICER-only data favors slightly softer EOS, see the previous discussion, the maximum-mass estimate decreases slightly to $M_{\rm max}=\MmaxBoundNICERRiley (\MmaxBoundNICERMiller)$.
When additionally considering the XMM data, the new data prefers slightly stiffer EOS but the maximum mass estimate, $\MmaxBoundNICERXMMRiley (\MmaxBoundNICERXMMMiller)$, does not change significantly.
The reason is that the upper limit on $M_{\rm max}$ is mainly determined by the constraint of \cite{Rezzolla:2017aly}.
Because of the strong impact of this constraint, we also consider the scenario where this upper limit on $M_{\rm max}$ is not included. 
In this case, $M_{\rm max}$ is found to be $2.34^{+0.34}_{-0.28}M_{\odot}$ without the new NICER data and changes to $\MmaxNoBoundNICERRiley (\MmaxNoBoundNICERMiller)$ including the NICER-only data and to $\MmaxNoBoundNICERXMMRiley (\MmaxNoBoundNICERXMMMiller)$ for the NICER and XMM data.
The changes are small because the increased stiffness coming from the pulsar radius measurement competes with the updated lower pulsar mass.

The posterior of $M_{\rm max}$ affects the classification of the secondary component of GW190814~\citep{Abbott:2020khf}, where a $2.6 M_{\odot}$ compact object merged with a $22 M_{\rm \odot}$ black hole.
Due to its extreme mass ratio and the low primary spin, the nature of the secondary component cannot be extracted from observational data. 
Instead, it has to be extracted from EOS modeling, see, e.g, \cite{Kalogera:1996ci}, \cite{Biswas:2020xna}, \cite{Essick:2020ghc}, and \cite{Tews:2020ylw}.
To examine the probability for the secondary component of GW190814 to be a NS, the posterior of $M_{\rm max}$, $p_{M_{\rm{max}}}(m)$ and that of GW190814's $m_2$, $p_{m_2}(m)$ are compared. 
The probability for GW190814 being a NS-black hole merger is given by~\citep{Tews:2020ylw},
\begin{equation}
    \begin{aligned}
    &P({\rm GW190814 \ is \ NSBH})\\
    &= \int_{0}^{\infty}d\Delta m \int_{-\infty}^{\infty} dm\, p_{M_{\rm{max}}}(m+\Delta m)p_{m_2}(m).
    \end{aligned}
\end{equation}
Due to the strong tension between GW190814's $m_2$ and the upper limit from \cite{Rezzolla:2017aly}, the inclusion of the NICER measurement of PSR J0740+6620 does not impact the classification of this system.
With or without the new NICER measurement, the probability for GW190814 to be NS-black hole merger is estimated to be $< 0.1\%$.
However, if we relax the upper limit on $M_{\rm max}$, i.e., do no include the analysis of \cite{Rezzolla:2017aly}, the probabilities change. 
Using the NICER-only data, the probability for GW190814 to be a NS-black hole system changes to 6.30(10.5)\%, lower than the previous estimation of 19\% in \cite{Tews:2020ylw}. 
When additionally including XMM data, the probability for GW190814 to be a NS-black hole system changes to 15.2 (24.4)\%.
The corresponding posterior distributions are shown in Fig.~\ref{fig:Mmax_PDF}. 
Given all current observational and theoretical knowledge of the NS EOS, a binary black hole merger remains the most consistent scenario for GW190814.

\subsection{Existence of a phase transition}~\label{sec:PT}

QCD predicts that nucleonic matter undergoes a phase transition to quark matter at very high densities. 
If such a phase transition is realized in neutron stars, at which exact density such a phase transition would occur, and which properties this phase transition would exhibit is unknown \citep{Glendenning:1992vb, Alford:2001mh, Alford:2006vz}.
A strong first-order phase transition, i.e., a segment in the EOS where the speed of sound vanishes, would be a 'smoking gun' signature for the existence of exotic forms of matter inside NSs. 

Here, we calculate the Bayes factor $\mathcal{B}^{\rm PT}_{\rm NPT}$ for the presence of such a strong first-order phase transition against the absence of it. The Bayes factor $\mathcal{B}^{\rm PT}_{\rm NPT}$ is given by
\begin{equation}
    \begin{aligned}
    \mathcal{B}^{\rm PT}_{\rm NPT} &\equiv \frac{P(d|\mathcal{H}_{\rm PT})}{P(d|\mathcal{H}_{\rm NPT})}\\
    &= \left .\frac{P(\mathcal{H}_{\rm PT}|d)}{P(\mathcal{H}_{\rm NPT}|d)} \middle/ \frac{P(\mathcal{H}_{\rm PT})}{P(\mathcal{H}_{\rm NPT})}, \right.
    \end{aligned}
\end{equation}
where $P(\mathcal{H}_{\rm PT}|d)$ ($P(\mathcal{H}_{\rm NPT}|d)$) is the posterior probability for the presence (absence) of phase transition, and $P(\mathcal{H}_{\rm PT})$ ($P(\mathcal{H}_{\rm NPT})$) is the corresponding prior probability. 
A Bayes factor larger than one indicates that the presence of a phase transition is preferred, while a Bayes factor smaller than one suggests that the presence of a phase transition is disfavoured. 
Without information from the NICER measurement of PSR J0740+6620, we find the Bayes factor to be $\BayesNMMA$. 

When including NICER-only data, softer EOS are preferred and the Bayes factors in favor of a phase transition change to $\BayesNICERRiley\ (\BayesNICERMiller)$.
In this case, the NICER radius measurement of PSR~J0740+6620 alone slightly increases the Bayes factor for the presence of a strong first-order phase transition within a NS with respect to the original NMMA analysis, but such a transition remains disfavoured considering all data.
On the other hand, with the additional inclusion of the XMM data, the Bayes factor changes to $\BayesNICERXMMRiley\ (\BayesNICERXMMMiller)$.
Following the interpretation of the Bayes factor described in \cite{Jeffreys61}, the presence of a phase transition inside a NS is disfavoured in all cases, yet it is not ruled-out, in agreement with the findings of \cite{Somasundaram:2021ljr}.

In all cases, even though a radius measurement of PSR J0740+6620 probes the EOS at the highest densities we can observe in the Cosmos to date, the NICER data adds only limited information due to its sizable uncertainties. 
In addition, it remains inconclusive if the radius measurement of PSR J0740+6620 itself suggests the presence of a phase transition inside a NS because the NICER-only and NICER-XMM data shift the Bayes factors in different directions; cf.~last column in Tab.~\ref{tab:results}. Moreover, an analysis using a different EOS prior shows that the direction of the shift is prior-dependent; cf.~ Tab.~\ref{tab:results_natural}.

\begin{table*}
\centering
\tabcolsep=0.2cm
\def\arraystretch{1.9}
\caption{Summary of the resulting posteriors for the radius of a typical NS $R_{1.4}$, the NS maximum mass $M_{\rm max}$, and the Bayes factor for phase transition against its absence, $\mathcal{B}^{\rm PT}_{\rm NPT }$. 
The values shown outside (inside) parentheses refer to the results without (with) inclusion of XMM data. 
All quoted errors are given at 90\% credible interval.} 
\label{tab:results}
\begin{tabular}{c||c|ccc}
\hline
\vspace*{-0.4cm}
 Quantity & NMMA        & NMMA                 & NMMA              & NMMA + Combined\\
          &             & + Miller et al. 2021 & + Riley et al. 2021 & Miller \& Riley et al. 2021 \\
\hline
$R_{1.4}$ & $11.75^{+0.86}_{-0.81}{\rm km}$ & $\ROnePointFourNICERMiller\ (\ROnePointFourNICERXMMMiller)$ & $\ROnePointFourNICERRiley\ (\ROnePointFourNICERXMMRiley)$ & $\ROnePointFourNICERBoth\ (\ROnePointFourNICERXMMBoth)$ \\
$M_{\rm max}$ & $2.18^{+0.14}_{-0.13}M_{\odot}$ & $\MmaxBoundNICERMiller\ (\MmaxBoundNICERXMMMiller)$ & $\MmaxBoundNICERRiley\ (\MmaxBoundNICERXMMRiley)$ & $\MmaxBoundNICERBoth\ (\MmaxBoundNICERXMMBoth)$\\
$\mathcal{B}^{\rm PT}_{\rm NPT }$ & $\BayesNMMA$ & $\BayesNICERMiller\ (\BayesNICERXMMMiller)$ & $\BayesNICERRiley\ (\BayesNICERXMMRiley)$ & $\BayesNICERBoth\ (\BayesNICERXMMBoth)$\\
\hline
\end{tabular}
\end{table*}


\section{Summary}~\label{sec:summary}
Using our nuclear physics -- multi-messenger astronomy framework~\citep{Dietrich:2020lps}, we have studied the impact of the new NICER observations of PSR J0740+6620 on the neutron-star EOS. 
While the NICER data alone shows good agreement with our previous NMMA predictions and therefore, validates our results, the additional inclusion of XMM data prefers slightly stiffer EOS.
However, due to the large uncertainties of 10-20\% in the NICER radius measurement of PSR J0740+6620, changes remain small.

In particular, the radius of a $1.4M_{\odot}$ NS $R_{1.4}$ changes from  $R_{1.4}=11.75^{+0.86}_{-0.81}$~km~\citep{Dietrich:2020lps} to $\ROnePointFourNICERRiley$ and $\ROnePointFourNICERMiller$ for the analyses from \cite{Riley:2021pdl} and \cite{Miller:2021qha}, respectively, at 90\% confidence without the XMM data and to $\ROnePointFourNICERXMMRiley$ and $\ROnePointFourNICERXMMMiller$ with the XMM data.
Combining the latter results, we obtain a final radius estimate of $\ROnePointFourNICERXMMBoth$ (at 90\% confidence), showing excellent agreement with our initial prediction.
Although the NICER-XMM data is informative, its large measurement uncertainties prevent it from significantly influencing our NMMA analysis. 

We also investigated its impact on the maximum allowed NS mass, $M_{\rm max}$, and its influence on the probability for GW190814 to be a NS-black hole merger. 
The upper limit on the maximum mass is mainly influenced by electromagnetic observations of GW170817~\citep{Margalit:2017dij,Rezzolla:2017aly} and therefore, the NICER data does not result in an observable impact.
When not enforcing this upper bound on $M_{\rm max}$, the probability for GW190814 to be a NS-black hole merger changes from 19\% to $6.3\%$ and $10.5\%$ ($15.2\%$ and $24.4\%$) with the inclusion of NICER (NICER+XMM) data from \cite{Riley:2021pdl} and \cite{Miller:2021qha}, respectively. 
Based on these estimations, it remains most plausible that GW190814 originated from a binary black-hole merger.

Finally, we studied the possibility for a first-order phase transition to be present inside NSs. 
Following the interpretation of Bayes factors suggested in \cite{Jeffreys61}, the presence of phase transition inside NSs is disfavoured, yet it is not ruled-out. 
However, this result is mainly impacted by previous multi-messenger observations of NSs and the impact of the new NICER measurement is small.

Observation of NSs have the potential to help us answer key questions in nuclear physics but current uncertainties of individual data remain large.
This highlights the importance of flexible multi-messenger frameworks that can use input from nuclear theory modeling of the EOS, laboratory experiments, and complementary observations of NSs to probe different aspects and to paint a complete picture of the EOS.

\section*{Acknowledgement}
We thank LIGO's extreme matter group for valuable comments and discussions. 
We are also grateful to the NICER collaboration for releasing their posterior samples and collected data. 
P.T.H.P and C.V.D.B are supported by the research program of the Netherlands Organization for Scientific Research (NWO). 
The work of I.T. was supported by the U.S. Department of Energy, Office of Science, Office of Nuclear Physics, under contract No.~DE-AC52-06NA25396, by the Laboratory Directed Research and Development program of Los Alamos National Laboratory under project numbers 20190617PRD1 and 20190021DR, and by the U.S. Department of Energy, Office of Science, Office of Advanced Scientific Computing Research, Scientific Discovery through Advanced Computing (SciDAC) program.
M.W.C. acknowledges support from the National Science Foundation with grant number PHY-2010970.
M.B. acknowledges support from the Swedish Research Council (Reg. no. 2020-03330).
Computations were performed on the national supercomputer Hawk at the High Performance Computing Center Stuttgart (HLRS) under the grant number 44189 and on SuperMUC-NG at Leibniz Supercomputing Centre Munich under project number pn29ba. 
Computational resources have also been provided by the Los Alamos National Laboratory Institutional Computing Program, which is supported by the U.S. Department of Energy National Nuclear Security Administration under Contract No.~89233218CNA000001, and by the National Energy Research Scientific Computing Center (NERSC), which is supported by the U.S. Department of Energy, Office of Science, under contract No.~DE-AC02-05CH11231.
This research has made use of data, software and/or web tools obtained from the Gravitational 
Wave Open Science Center (https://www.gw-openscience.org), a service of LIGO Laboratory, the 
LIGO Scientific Collaboration and the Virgo Collaboration. LIGO is funded by the U.S. National 
Science Foundation. Virgo is funded by the French Centre National de Recherche Scientifique (CNRS), 
the Italian Istituto Nazionale della Fisica Nucleare (INFN) and the Dutch Nikhef, with contributions 
by Polish and Hungarian institutes.

\bibliography{main}

\appendix
\section{Impact of EOS prior}
\begin{table*}[h]
\centering
\tabcolsep=0.2cm
\def\arraystretch{1.9}
\caption{Summary of the resulting posteriors without an uniform prior on $R_{1.4}$ imposed, for the radius of a typical NS $R_{1.4}$, the NS maximum mass $M_{\rm max}$, and the Bayes factor for phase transition against its absence, $\mathcal{B}^{\rm PT}_{\rm NPT }$. 
The values shown outside (inside) parentheses refer to the results without (with) inclusion of XMM data. 
All quoted errors are given at 90\% credible interval.} 
\label{tab:results_natural}
\begin{tabular}{c||c|ccc}
\hline
\vspace*{-0.4cm}
 Quantity & NMMA        & NMMA                 & NMMA              & NMMA + Combined\\
          &             & + Miller et al. 2021 & + Riley et al. 2021 & Miller \& Riley et al. 2021 \\
\hline
$R_{1.4}$ & $11.94^{+0.58}_{-0.64}{\rm km}$ & $11.86^{+0.58}_{-0.58}{\rm km} (12.04^{+0.61}_{-0.61}{\rm km})$ & $11.82^{+0.56}_{-0.57}{\rm km} (11.96^{+0.62}_{-0.57}{\rm km})$ &  $11.84^{+0.56}_{-0.59}{\rm km} (12.00^{+0.61}_{-0.60}{\rm km})$\\
$M_{\rm max}$ & $2.19^{+0.14}_{-0.13}M_{\odot}$ & $2.18^{+0.15}_{-0.12}M_{\odot} (2.18^{+0.18}_{-0.13}M_{\odot})$ & $2.17^{+0.15}_{-0.11}M_{\odot} (2.17^{+0.16}_{-0.12}M_{\odot})$ & $2.18^{+0.14}_{-0.12}M_{\odot} (2.18^{+0.17}_{-0.13}M_{\odot})$ \\
$\mathcal{B}^{\rm PT}_{\rm NPT }$ & $0.64 \pm 0.01$ & $0.62 \pm 0.01 (0.67 \pm 0.01)$ & $0.60 \pm 0.01 (0.64 \pm 0.01)$ & $0.61 \pm 0.01 (0.65 \pm 0.01)$\\
\hline
\end{tabular}
\end{table*}

In Table~\ref{tab:results_natural}, we present the summary of the resulting posteriors for the quantities of interest using an EOS prior that is not uniform in $R_{1.4}$, i.e., a prior that is ``natural" to our speed-of-sound extension scheme. 
The posteriors on $R_{1.4}$ and $M_{\rm max}$ are consistent with the results shown in Tab.~\ref{tab:results} within the uncertainties.

In contrast, the Bayes factors $\mathcal{B}^{\rm PT}_{\rm NPT}$ are prior-sensitive, and their values shift significantly from the results in Tab.~\ref{tab:results}. 
Because the non-uniform prior set does not explore as many EOS with phase transitions as the uniform $R_{1.4}$ prior set, the shifts of the Bayes factors changes. 
However, the presence of a phase transition inside a NS is still disfavoured in all cases, but cannot be ruled-out.

\end{document}